\documentstyle[preprint,aps]{revtex}
\begin{document}

\draft
\begin{title}
{A Mixed Basis Approach for the Efficient Calculation
of Potential Energy Surfaces}
\end{title}

\author{O. G\"{u}lseren, D.M. Bird and S.E.~Humphreys}
\address{School of Physics, University of Bath,
Claverton Down, Bath BA2 7AY, UK.}

\date{\today}
\maketitle

\begin{abstract}

First principles calculations based on density functional theory are
having an increasing impact on our understanding of molecule--surface
interactions. For example, calculations of the multi-dimensional 
potential energy surface have provided considerable insight into the 
dynamics of dissociation processes. However, these calculations using
a plane-wave basis set are very compute expensive if they are to be fully
converged with respect to the plane-wave energy cutoff, k--point sampling,
supercell size, slab thickness, etc. Because of this, in this study, we
have implemented a mixed-basis approach which uses pseudo-atomic orbitals
and a few low-energy plane waves as the basis set within a density
functional, pseudopotential calculation. We show that the method offers a
computationally cheap but accurate alternative. The energy barrier for 
hydrogen dissociation on Cu(111) is calculated as an example.

\end{abstract}

\pacs
{ \\
Keywords: Chemisorption, Copper, Density functional calculations,
Ab-initio quantum chemical methods and calculations, Hydrogen,
Low index single crystal surfaces, Models of surface chemical reactions}

\section{Introduction and Computational Method}

In recent years, the application of first principles electronic structure
methods to surface systems has increased significantly as a result
of improvements in algorithms and enhanced computational speed. The system 
size and complexity which can be analysed has increased by an order of 
magnitude following the pioneering work of Car and Parrinello\cite{carpar},
who introduced an iterative minimisation of the total energy based on
wavefunction improvements at each iteration. The construction of the initial 
wavefunction is clearly important for efficiency in such an approach.

Among several different approaches, there are two simple and natural choices
of basis set for the expansion of electron wavefunctions: atomic orbitals and
plane waves. On the negative side, atomic orbital methods have difficulties
in representing the wavefunctions and potential in interstitial and vacuum
regions while plane wave expansions are expensive for representing 
localised atomic character, for example 3d wavefunctions. Nevertheless, 
plane-wave 
basis sets are in most common use since they are simple, independent of 
atomic positions, fast Fourier transformation (FFT) methods can be applied 
readily, and accuracy can be systematically improved by including 
additional plane waves with higher energy cut-offs. Although atomic 
orbitals are more physical it is difficult to represent uniform charge 
density, as in the vacuum region of a surface, with atom-centered, 
localised orbitals. On the other hand, plane-wave basis sets are also 
inefficient in a surface calculation using a slab geometry, since one 
needs as many plane waves for the vacuum as for the solid region.
Therefore, a combination of the important properties of plane waves with
atomic orbitals in a mixed basis may give a convenient and efficient
representation, especially for systems which include both highly localised
(atomic-like) and delocalised (plane-wave-like) components.

In this study, we have implemented a mixed-basis approach~\cite{mixed}
which uses pseudo-atomic orbitals and a few low-energy plane waves as the
basis set within a density functional, pseudopotential calculation. A
similar approach has been described by Neugebauer and Van de Walle 
\cite{initial},
but they focussed on providing a better starting wavefunction for a full
plane-wave calculation. Here, we investigate whether the mixed-basis
calculation might be used in its own right for calculations of the potential
energy surface for molecule-surface systems.

The Kohn-Sham eigenfunctions are expanded as
\begin{equation}
\psi_{\alpha\vec{k}}(\vec{r}) = 
\sum_{\mu} a_{\mu}^{\alpha}(\vec{k}) \chi_{\mu}(\vec{r}) +
\frac{1}{\sqrt{\Omega}} \sum_{\vec{G}} b_{\vec{G}}^{\alpha}
e^{i(\vec{k}+\vec{G}).\vec{r}}
\label{eigfun}
\end{equation} 
where $\alpha$ is the band index, $\mu$ is a combined index which
labels the orbitals and atomic sites, $a$ and $b$ are coefficients of the 
pseudo-atomic orbitals and plane waves, respectively, and $\Omega$ is the
volume of the unit cell. $\chi_{\mu}$ is the Bloch sum formed from 
pseudo-atomic orbitals as
\begin{equation}
\chi_{\mu}(\vec{r}) \equiv \chi_{m}^{i}(\vec{r}) = 
\sum_{\vec{R_l}} e^{i\vec{k}.(\vec{R_l}+\vec{\tau_i})}
\phi_m(\vec{r}-\vec{R_l}-\vec{\tau_i})
\end{equation} 
where $m$ labels the orbitals, the $\vec{R_l}$ are the lattice vectors,
the $\vec{\tau_i}$ are the atomic coordinates, and $\phi_{m}$ are 
pseudo-atomic orbitals. In practice, we use a plane-wave expansion for
$\chi_{\mu}(\vec{r})$, and exactly the same FFT grid as in a full 
plane-wave calculation. There are therefore two plane-wave energy cut-offs
to be considered in the mixed-basis calculation. The larger one is the 
cut-off used in the representation of $\chi_{\mu}(\vec{r})$ and is the 
same as would be used in a full plane-wave calculation. The smaller one
is the cut-off for the {\it extra}, low-energy plane waves which appear
in the second term of Eq.~\ref{eigfun}. This plane-wave representation of 
$\chi_{\mu}(\vec{r})$ makes the calculation of the charge density, 
the kinetic energy, multicentre integrals, and the contribution
from non-local pseudopotentials straightforward.

Solving the Schr\"{o}dinger equation then reduces to solving the secular 
equation
\begin{equation}
det | H - SE | = 0.
\label{secular}
\end{equation}
The overlap matrix elements are given by (with reference to the
partition of $\psi$ in Eq.\ref{eigfun})
\begin{eqnarray}
S_{\vec{G}\vec{G'}} & = &  \delta_{\vec{G},\vec{G'}} \\
S_{\mu\vec{G}} & = & e^{-i\vec{G}.\vec{\tau_{i}}}
I_{\vec{G}}^{m}(\vec{k}) \\
S_{\mu\nu} & = & \sum_{\vec{g}} 
e^{-i\vec{g}.(\vec{\tau_{i}}-\vec{\tau_{j}})}
I_{\vec{g}}^{n\star}(\vec{k})
I_{\vec{g}}^{m}(\vec{k})
\end{eqnarray}
where $\nu\equiv(n,j)$ and $I_{\vec{g}}^{m}(\vec{k})$ is the 
Fourier integral of the pseudo-atomic orbital,
\begin{equation}
I_{\vec{g}}^{m}(\vec{k}) =
\frac{1}{\sqrt{\Omega}} \int d\vec{r} e^{-i(\vec{k}+\vec{g}).\vec{r}}
\phi_{m}(\vec{r}).
\end{equation}
Similarly, the Hamiltonian matrix elements are based on a plane-wave 
representation
\begin{eqnarray}
H_{\vec{G}\vec{G'}} & = &
\frac{1}{\sqrt{\Omega}} \int d\vec{r} 
e^{-i(\vec{k}+\vec{G'}).\vec{r}}
\hat{H}
e^{i(\vec{k}+\vec{G}).\vec{r}} \\
 & = & |\vec{k}+\vec{G}|^2 \delta_{\vec{G},\vec{G'}}
+ V_{local}(\vec{G}-\vec{G'})
+ V_{NL}((\vec{k}+\vec{G}),(\vec{k}+\vec{G'})).
\label{hamg}
\end{eqnarray}
Then
\begin{eqnarray}
H_{\mu\vec{G}} & = &  \sum_{\vec{g}}
e^{-i\vec{g}.\vec{\tau_{i}}}
I_{\vec{g}}^{m}(\vec{k}) H_{\vec{g}\vec{G}} \\
H_{\mu\nu} & = &  \sum_{\vec{g}\vec{g'}} 
e^{i\vec{g'}.\vec{\tau_{j}}}
e^{-i\vec{g}.\vec{\tau_{i}}}
I_{\vec{g'}}^{n\star}(\vec{k})
I_{\vec{g}}^{m}(\vec{k}) H_{\vec{g}\vec{g'}}
\end{eqnarray}
The local part of the potential, $V_{local}$, in Eq.~\ref{hamg} contains
the Hartree and exchange-correlation potentials, as well as the local part
of the pseudopotential. In practice, only the Hartree and exchange-correlation 
contributions need be re-calculated through the self-consistency cycle---the
pseudopotential (both local and non-local, $V_{NL}$, parts) and kinetic 
energy matrix
elements are calculated only at the first iteration. Self-consistency is
achieved by a combination of Kerker charge density mixing and a
modified Broyden method \cite{mixing,kresse}. The initial charge density
is constructed from overlapping, atomic pseudo-charge densities.

Diagonalization of Eq.~\ref{secular} is acceptable, since there are at most
9 orbitals (s,p,d) for each atom, and typically 10 to 20 additional plane
waves per atom (see below). This results in a matrix size less than
$10^3\times10^3$, even for a moderately large system, compared to between
$10^4$ to $10^5$ for a pure plane-wave expansion. In fact, for the 
H$_2$/Cu system considered below, the most expensive part of the whole 
calculation is the construction of the Hamiltonian matrix. Tests on this
system show that the mixed-basis method is 6 to 8 times faster per 
iteration than our pure plane wave code (as described in~\cite{casimp}).
In addition, it typically requires fewer than half as many
iterations to converge, and so provides a significant improvement in
computational speed. A full analysis of the timing and scaling of the
computation with respect to the important calculational parameters will
be presented elsewhere.

\section{Calculations and Results}

We have carried out a careful benchmarking analysis of the mixed-basis
approach by increasing the number of additional low-energy plane-waves 
and comparing the results with an exactly equivalent calculation based 
on plane waves only. The system chosen for this comparison is hydrogen
dissociation on Cu(111), which is a model system for the experimental
(eg~\cite{cuexp}) and theoretical (eg~\cite{h2cuprl,gross,h2cuss}) study of
dissociation dynamics. In particular, we concentrate on the value of the
minimum energy barrier, which is known to occur for the geometry in which
the molecular axis is parallel to the surface plane and with the H$_2$
molecule over a bridge site with the H atoms pointing towards neighbouring
hollow sites~\cite{h2cuprl}.
In all calculations the transition state is taken to be where the H$_2$
molecule is 1.2\AA\ above the top-layer Cu atoms, with a bond length of
1.1\AA~\cite{h2cuss}. 

As a preliminary study, we have examined some properties of bulk fcc
Cu and the H$_2$ molecule within the mixed-basis scheme. A
semi-relativistic, Troullier-Martins\cite{psepot} pseudopotential
(with associated pseudo-atomic orbitals) is used to describe the Cu
atoms, and hydrogen is described by the full Coulombic potential, with
localised 1s and 2p orbitals. For Cu, the irreducible wedge of the fcc 
Brillouin zone is sampled with 28 $\vec{k}$--points and the H$_2$ 
molecule is calculated in the same cell as in the H$_2$/Cu(111) system
described below. The calculated lattice parameter and bulk modulus
for Cu; and bond length and vibrational frequency (estimated from a
harmonic fit about the equilibrium bond length) for H$_2$ are presented
in Tables I and II respectively, as a function of the cut-off energy for the 
low-energy plane waves. It can be seen that for these two very different
systems the mixed-basis method converges rapidly to the full plane-wave 
result. 

Technical details of the full H$_2$/Cu(111) calculation are as follows. The 
substrate is modelled by a 5 layer, rigid Cu slab with the experimental
lattice parameter of 3.61\AA. There are 3 Cu atoms per layer within 
a $\sqrt{3}\times\sqrt{3}$ geometry and 4 layers of vacuum separate the
slabs. An H$_2$ molecule is placed on only one side of the slab. The 
surface Brillouin zone is sampled by 54 special $\vec{k}$-points 
(15 in the irreducible wedge), and the Fermi surface is broadened with a 
0.25eV smearing function, with the total energy being extrapolated to zero 
temperature. A cut-off of 800eV is used for the pure plane-wave calculation, 
and for the representing the localised orbitals in the mixed-basis method. 
These calculational parameters should provide well-converged results 
for the energy barrier~\cite{h2cuss}. The local density approximation (LDA) 
is used in the self-consistency cycle of the calculations, and the
generalised gradient approximation (GGA) (see eg~\cite{gracor}) is 
incorporated by calculating the exchange-correlation energy of the LDA 
density.

Six values of the minimum barrier have been calculated. Five use the 
mixed-basis approach with different numbers of additional plane waves,
corresponding to energy cut-offs of 0eV (ie pure pseudo-atomic orbitals), 
20eV, 40eV, 60eV and 80eV. 
The typical number of additional plane waves are 
0, 65, 180, 330 and 520 respectively for these cut-offs.  
The sixth value corresponds to the full 
plane-wave calculation. 
Computed values of the minimum energy barrier are given in 
Table~\ref{ener:barrier}. It can be seen that the mixed-basis method
provides an accurate value for the barrier height (within 10 meV) once 
plane waves with energies up to 40 eV are included.

\section{Conclusions}

The mixed-basis method we have presented appears to provide an accurate
and computationally cheap alternative to full plane-wave methods for 
first principles calculations of surface systems. Although we have
discussed only a single case in detail, tests on a variety of systems
have confirmed the generality of this approach, with structural
parameters being accurate to of order 0.01\AA\ and energy differences 
accurate to a few tens of meV. Only total energies have been discussed
here, but it is also straightforward to calculate forces within the
mixed-basis scheme using the general force expressions given in
\cite{force}. This holds out the prospect of being able to perform, with
high computational efficiency, a full structural relaxation of a system 
using the
mixed-basis approach. If required, the calculation can then be ``finished
off'' using a full plane-wave expansion, in which case the mixed-basis
method will provide an accurate starting structure and highly optimised 
initial wavefunctions.

This work is supported by the UK Engineering and Physical Sciences Research
Council.

\clearpage

\begin{table}
\begin{tabular}{c c c}
Energy cut-off (eV) & Lattice constant (\AA) & Bulk Modulus (GPa) \\
\hline
AO & 3.59 & 165 \\
20 & 3.57 & 180 \\
40 & 3.57 & 177 \\
60 & 3.58 & 173 \\
80 & 3.58 & 175 \\
PW & 3.57 & 178 \\
\end{tabular}
\vspace*{0.5cm}
\caption{Calculated LDA lattice constant and bulk modulus for fcc Cu, as
a function of the cut-off energy for additional plane waves in the
mixed-basis approach. AO represents the pure pseudo-atomic orbital
limit (ie 0 eV cut-off) and PW the full plane-wave limit (with plane
waves up to 800 eV). The Murnaghan equation of state is used.}
\label{bulk}
\end{table}

\begin{table}
\begin{tabular}{c c c}
Energy cut-off (eV) & Bond length (\AA) & Vibrational quantum (meV) \\
\hline
AO & 0.837 & 535 \\
20 & 0.781 & 489 \\
40 & 0.761 & 471 \\
60 & 0.756 & 467 \\
80 & 0.754 & 464 \\
PW & 0.753 & 462 \\
\end{tabular}
\vspace*{0.5cm}
\caption{Calculated GGA bond length and vibrational quantum ($\hbar\omega$) 
for H$_2$. Symbols are as in Table I.}
\label{h2dimer}
\end{table}

\begin{table}
\begin{tabular}{c c}
Energy cut-off (eV) & Barrier Height (eV) \\
\hline
AO & -0.854 \\
20 &  0.520 \\
40 &  0.606 \\
60 &  0.614 \\
80 &  0.620 \\
PW &  0.612 \\
\end{tabular}
\vspace*{0.5cm}
\caption{GGA minimum energy barrier for H$_2$ dissociation on Cu(111).
Symbols are as in Table I.}
\label{ener:barrier}
\end{table}

\end{document}